\documentclass[aps,prb,twocolumn,showpacs,groupedaddress,floatfix]{revtex4}
\usepackage{graphicx}% Include figure files
\begin{document}

%\draft
%\twocolumn[    %+++++
%\hsize\textwidth\columnwidth\hsize\csname @twocolumnfalse\endcsname    %+++++

\title{ Unconventional superconducting pairing symmetry
induced by phonons} 

\author{I. Schnell,$^1$ I. I. Mazin,$^{1,2}$ and Amy Y. Liu$^1$}
\affiliation{$^1$ Department of Physics, Georgetown University, 
  Washington, DC 20057-0995 \\
$^2$ Center for Computational Materials Science,
Naval Research Laboratory, Washington, DC 20375} 
\date{\today}
%\maketitle

%\widetext
\begin{abstract}
The possibility of non-$s$-wave superconductivity induced
 by phonons is investigated using a simple model that is inspired 
by Sr$_2$RuO$_4$.  The model assumes a two-dimensional  electronic
structure, a two-dimensional spin-fluctuation spectrum,
and three-dimensional electron-phonon coupling.
Taken separately, each interaction 
favors formation of spin-singlet pairs (of $s$ symmetry for the phonon
interaction and $d_{x^2-y^2}$ symmetry for the spin interaction), 
but in combination, 
a variety of more unusual singlet and triplet states are found, 
depending on the interaction parameters. 
This may have important implications for Sr$_2$RuO$_4$,
providing a plausible explanation of how the observed
spin fluctuations, which clearly favor  $d_{x^2-y^2}$
pairing, may still be instrumental in creating a superconducting state 
with a different ($e.g.,$ $p$-wave) symmetry. It also suggests 
an interpretation of the large isotope effect observed in
Sr$_2$RuO$_4$. These results indicate that phonons could play a key role
in establishing the order-parameter symmetry in Sr$_2$RuO$_4$,
and possibly in other unconventional superconductors.

\end{abstract}

      %+++++

%\narrowtext
%============================================================================
\maketitle
\section{Introduction}

The existence of superconducting states that break symmetries of the 
normal state beyond gauge symmetry has generated considerable interest. 
Such unconventional pairing was first 
discovered in $^3$He, in which the three superfluid phases all have 
spin-triplet ($S=1$) atomic Cooper pairs with relative orbital angular 
momentum $L=1$ ($p$-wave).\cite{Leggett}  
Strong evidence for unconventional superconductivity 
was subsequently found in heavy-fermion materials.\cite{heavyfermions}
In the high-$T_c$ cuprates, phase-sensitive experiments have
provided convincing evidence for $L = 2$ ($d$-wave) spin-singlet
pairing.\cite{Tsuei}  The pairing symmetry in Sr$_2$RuO$_4$, 
a non-cuprate perovskite oxide superconductor, has been
unambiguously shown to be lower than the symmetry of the underlying
crystal lattice, {\it i.e.}, $L > 0$.\cite{Mackenzie}
It is generally believed that Sr$_2$RuO$_4$ is a triplet $p$-wave 
superconductor,\cite{Mackenzie} though it has been noted that
a $d$-wave singlet state is also possible.\cite{Zutic}
It has been suggested that some compounds could be
superconductors with $L > 2$.\cite{Sauls,Johannes} More
exotic types of pairing, such as order parameters with odd-frequency 
dependence and mixed singlet-triplet Cooper pairs,
have also been considered for real materials.\cite{Johannes,Bergeret,Yuan} 

By itself, the electron-phonon interaction is always attractive, and hence
it maximizes the pairing energy for an order parameter that is positive
everywhere on the Fermi surface.  Conventional wisdom thus suggests
that $L > 0$ pairing symmetries must be induced by
non-phonon mechanisms, such  as magnetic or Coulomb interactions. 
A number of authors have pointed out, however, that
while phonons are always more pairing in the $s$ channel,
they may also be pairing in other channels as well.\cite{phonons}  
If, for example,
strong on-site Coulomb repulsion suppresses $s$-wave pairing because
of the large overlap of the wave functions of the paired particles, or
if there is an additional pairing contribution in an $L>0$ channel
from other (non-phonon) mechanisms, unconventional pairing that
results mainly from electron-phonon coupling is possible. 
Less well appreciated is the idea that a combination of 
electron-phonon coupling and another pairing interaction
({\it e.g.}, spin-fluctuation exchange) may result in a superconducting
state that is not the ground state of either of the two
interactions taken separately. 

In this paper we consider  a simple model, loosely based on
Sr$_2$RuO$_4$, that illustrates this effect. The model 
assumes a highly two-dimensional electronic structure,
and  it includes a two-dimensional spin-fluctuation-induced interaction 
that favors $d$-wave pairing of $x^2-y^2$  (or $\cos2\phi$) 
symmetry,\cite{notation}
and a three-dimensional electron-phonon interaction
that favors an $s$-wave state.  
The spin interaction is based on the experimentally observed spin-fluctuation 
spectrum of Sr$_2$RuO$_4$, which has negligible dispersion along  
$z$.\cite{spinfluctuation,sfz}
While less is known about the electron-phonon
coupling in Sr$_2$RuO$_4$, the three-dimensional nature of the crystal
structure suggests that the electron-phonon interaction in the
material may be much more three dimensional than the electronic 
structure itself.
When the coupling strengths of the two interactions are comparable,
we find that the symmetry of the emerging ground state
can be changed by tuning the widths of the interactions. 
Among the states that are found to be stable in certain regions of phase
space are  two-dimensional ($z$-independent) $p$-wave states and  
three-dimensional ($z$-dependent) singlet and triplet   states, all
of which arise from the interplay between the two interactions. 

\section{Model}

The spin dependence of the pairing in a superconductor can be taken into 
account using  a $2 \times 2$ matrix $\hat{\Delta}({\bf k})$
to represent the gap function.
Within this spin matrix formalism, the BCS gap equation has the form
\begin{equation}
  \Delta_{s_1 s_2}({\bf k}) = \sum_{{\bf k^\prime} s_3 s_4}
           V_{s_2 s_1 s_3 s_4} ({\bf k},{\bf k^\prime})
  \Delta_{s_3 s_4}({\bf k^\prime}) F({\bf k^\prime},  T),
  \label{eq: gapeq}
\end{equation}
where
\begin{equation}
  V_{s_1 s_2 s_3 s_4}({\bf k},{\bf k^\prime}) =
      \langle -{\bf k} s_1; {\bf k} s_2 | \hat{V} | -{\bf k^\prime} s_4;
  {\bf k^\prime} s_3 \rangle
\end{equation}
is the matrix element of the effective electron-electron 
pairing interaction $\hat{V}$, for which we adopt a
convention in which positive $V$ corresponds to an attractive interaction,
and $F$ is a function of the temperature $T$ and
the quasiparticle energy $E_{\bf k^\prime}$.\cite{Sigrist}
In a crystal with inversion symmetry, the 
pairing potential can be separated into spin-triplet (+)  and spin-singlet
(-) channels, 
\begin{equation}
V_{s_1 s_2 s_3 s_4}({\bf k},{\bf k^\prime}) = 
        V^{(+)}({\bf k},{\bf k^\prime})S^{(+)}_{s_1 s_2 s_3 s_4}  
         + V^{(-)}({\bf k},{\bf k^\prime})S^{(-)}_{s_1 s_2 s_3 s_4}, 
\end{equation}
where  
\begin{equation}
S^{(\pm)}_{s_1 s_2 s_3 s_4} = (\delta_{s_1 s_4} \delta_{s_2 s_3} \pm 
                             \delta_{s_1 s_3} \delta_{s_2 s_4} )/2.
\end{equation}
The gap equation decouples into separate equations for
the singlet and triplet channels. 
For singlet pairing, the gap matrix has the form
$\Delta_0({\bf k})i\hat{\sigma}_y$, 
while for triplet pairing, 
it has the form $[{\bf \hat{\sigma}}\cdot{\bf d}({\bf k})] i\hat{\sigma}_y$, 
where $\hat{\sigma}$ are Pauli spin matrices.  Since the total
wave function must be antisymmetric under interchange of particles,
the scalar order parameter $\Delta_0({\bf k})$
for singlet pairing must be even in ${\bf k}$ and 
the vector order parameter ${\bf d}({\bf k})$ must have odd parity.

 Near  $T=T_c$, the gap function is small and the gap equation can be 
linearized and reformulated as an eigenvalue problem: 
\begin{equation}
  f(T_c)\Delta_{s_1 s_2}({\bf k}) = \sum_{{\bf k^\prime} s_3 s_4}
  V_{s_2 s_1 s_3 s_4} ({\bf k},{\bf k^\prime})
  \Delta_{s_3 s_4}({\bf k^\prime}).
  \label{eq: gapeq2}
\end{equation}
In the singlet channel, this becomes an eigenvalue problem
for $V^{(-)}({\bf k},{\bf k^\prime})$, with eigenfunctions given by the
scalar order parameter $\Delta_0({\bf k})$. 
For triplet pairing, the eigenvectors of
$V^{(+)}({\bf k},{\bf k^\prime})$ correspond to components of the 
vector order parameter
${\bf d}({\bf k})$. 
Among eigenfunctions that have spatial symmetry compatible with
the spin pairing,  the one with the largest eigenvalue $f(T_c)$ 
produces the first superconducting instability that manifests as 
temperature is lowered, and hence defines the physical $T_c$.  

In our model, the pairing potential has two contributions: 
an electron-phonon interaction $V_{ph}$ that is always attractive
regardless of the symmetry of the spin pairing, and
an interaction $V_{sp}$ that is mediated by antiferromagnetic
spin fluctuations.  The latter is attractive in the 
triplet spin channel but repulsive in the singlet spin channel. 
Therefore, the total pairing interaction is $V = (V_{ph} + V_{sp})S^{(+)}
+ (V_{ph} - V_{sp})S^{(-)}$. 

The model system considered here is inspired by Sr$_2$RuO$_4$. 
A tetragonal, three-dimensional crystal structure with $D_{4h}$
symmetry is assumed, with lattice constants $a$ and $c$. 
The electronic structure is highly two-dimensional, with a cylindrical
Fermi surface of radius $k_F^\parallel = 0.9 \pi/a$, similar to  
the $\gamma$ sheet of the Fermi surface in Sr$_2$RuO$_4$.
The electron-phonon interaction is assumed to be three dimensional
in nature while the antiferromagnetic-spin-fluctuation-induced interaction
is two dimensional.
Both interactions depend only on 
${\bf q} = {\bf k}-{\bf k^\prime}$.  The spin-fluctuation-induced
interaction $V_{sp}({\bf q})$ is taken to be an isotropic
Gaussian  in $q_x$ and $q_y$, centered on ${\bf q} = (2\pi/3a, 2\pi/3a, q_z)$ 
and equivalent
lines in the Brillouin zone [{\it i.e.}, $(2\pi/3a, 4\pi/3a, q_z)$, 
$(4\pi/3a, 2\pi/3a, q_z)$, $(4\pi/3a, 4\pi/3a, q_z)$], in
accordance with both theory\cite{MS} and experiment.\cite{spinfluctuation} 
The electron-phonon interaction $V_{ph}(\bf{q})$  is
a three-dimensional Gaussian function centered   at
$(\pi/a,\pi/a,\pi/c)$, {\it i.e.}, the corners of the Brillouin zone.
 The electron-phonon interaction is 
allowed to be anisotropic, with a different width in the $z$ direction 
than in the plane.  The anisotropy of the electron-phonon interaction is
characterized by the parameter $\alpha = aw_{ph}^\parallel/cw_{ph}^z$,
where the widths are defined in terms of the Gaussian form $\exp[-(q/w)^2]$.

Because the Fermi surface in the model is perfectly cylindrical
and the interaction potentials are  separable in Cartesian coordinates,
the eigenvalue problem separates into $z$ and planar 
parts, with product eigenfunctions. The in-plane eigenfunctions
are of the form $\cos(m\phi)$ and $\sin(m\phi)$, while the $z$ 
eigenfunctions are of the form $\cos(nk_zc)$ and $\sin(nk_zc)$, 
where $m$ and $n$ are integers and $\tan\phi = k_y/k_x$.  
For $n \neq 0$, the two-dimensional spin interaction
cancels out since $\sum_{\bf k^\prime} V_{sp}({\bf k},{\bf k^\prime})
\cos(nk_z^\prime c)  =  \sum_{\bf k^\prime} V_{sp}({\bf k},{\bf k^\prime})
\sin(nk_z^\prime c) = 0$ if $V_{sp}$ has no
$z$ dependence.  This means there are degenerate spin-triplet and 
spin-singlet states that have the same in-plane eigenfunction.  The
symmetry of the in-plane eigenfunction determines which one
has the cosine solution and which one has the sine solution in $k_z$. 
In a real material, this degeneracy between triplet and
singlet states would be broken by a variety of effects,
including warping of the Fermi surface and the spin-orbit interaction.

Numerical calculations were carried out using a discrete grid of 
${\bf k}$ points on the Fermi surface. The circumference of the
Fermi surface was sampled with 40 equally spaced grid points, while
20 points were used along $k_z$. The grid was selected so that 
the spacing between  grid points was small compared to the
the widths of the interactions.
We have also checked convergence 
by increasing the number of grid points to $60\times 30$ and 
$80 \times 40$.

The effective strength of each interaction in coupling electrons on the
Fermi surface is measured by
\begin{equation}
  \lambda_{sp(ph)} = \sum_{\bf{k}, \bf{k^\prime} \in FS} 
  V_{sp(ph)}({\bf k},{\bf k^\prime}).
  %\delta(\epsilon_k - E_F)\delta(\epsilon_{k^\prime}-E_F).
\end{equation}
For fixed ratios of $R = \lambda_{ph}/\lambda_{sp}$, 
phase diagrams were mapped out 
as a function of interaction widths.  
Because of the discrete 
sampling of {\bf k}-points, results are presented only for interactions 
with Gaussian width greater that $0.1\pi/a$ in the plane. 
To construct the phase diagrams, we used a recursive algorithm that 
takes advantage of the fact that most phases occupy large regions in 
phase space. The gap equation is solved for all points along the border
of the rectangular phase diagram. If the ground state is the
same at all points on the border, all points in the interior are assumed
to have the same ground state.  Otherwise, the region is  divided into
two rectangles and the process is repeated recursively. 
For each phase diagram, phase space was sampled with a range of 
resolutions, from $5\times 5$ to $400\times 400$ points, to
mitigate the chance of missing phases that might be stable
in only small interior regions of the diagram. 

\section{Results}

We consider first the case when one of the 
interactions dominates.  When the spin coupling is 
much larger than the phonon coupling ($R \ll 1$), only
one phase appears regardless of the widths of the spin and 
phonon interactions.    This is the singlet $\cos{2\phi}$
state (commonly denoted $k_x^2 - k_y^2$).
Because the spin interaction is repulsive in the singlet channel,
to induce pair formation, it must scatter electrons between 
points {\bf k} and {\bf k$^\prime$} on the Fermi surface that have 
order parameter of opposite sign (see Eq.  \ref{eq: gapeq2}).
For $\cos{2\phi}$ symmetry, three of the four lines in the
Brillouin zone on which the spin interaction is maximal do this; 
the fourth couples nodal regions. Hence this
$d$-wave state is stabilized by the antiferromagnetic
spin fluctuations. When the phonon interaction dominates
($R \agt 5$), only an $s$-wave spin singlet appears in the phase
diagram.  This is expected since the attractive electron-phonon interaction
is pair forming when the order parameter has the same sign at points 
{\bf k} and {\bf k$^\prime$} on the Fermi surface. 

\begin{figure}[bf]
\includegraphics[width=2.7 in]{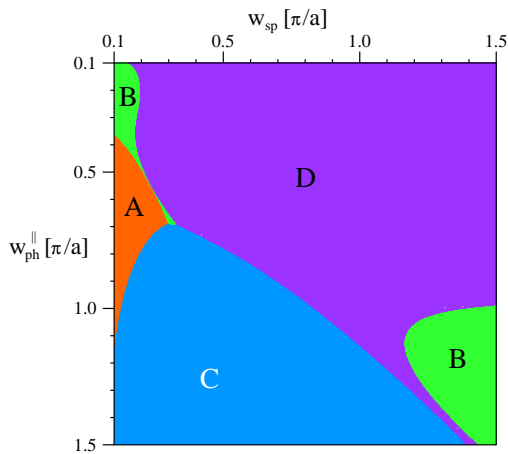}
\caption{Phase diagram for coupling-strength ratio
$R = \lambda_{ph}/\lambda_{sp} = 1$ and electron-phonon anisotropy $\alpha = 1$.
See Table \ref{tab: states} for the symmetry of states that
appear in this phase diagram.
\label{fig: phase1}}
\end{figure}

Figure \ref{fig: phase1}  
shows the phase diagram for equal spin and phonon coupling 
strengths ($R=1$).  
In the diagram, the width of the spin-fluctuation-induced
interaction increases from left to right, and the width of the electron-phonon
interaction increases from top to bottom.
The anisotropy parameter for the width of the electron-phonon interaction  
is $\alpha = 1$. 
In the regions labeled  B (green),
the $s$-wave singlet favored by the electron-phonon interaction is stable,
while in region C (blue), the $\cos{2\phi}$ spin-singlet state
favored by the spin-fluctuation-induced interaction is stable. 
New phases also appear.  There is 
a manifold of degenerate triplet states with  in-plane $p$-wave pairing 
(A, orange), and degenerate singlet and triplet states 
that have $k_z$-dependent order parameters (D, purple).     The
character of these phases is summarized in Table \ref{tab: states}.
We have mapped out phase diagrams for coupling-strength ratios ranging from 
$R=0$ to $R=5$.  With the electron-phonon anisotropy fixed at 
$\alpha = 1$, no states  other than the ones that are shown in the 
$R=1$ diagram are found to be stable.

\begin{table}[tbf]
\caption{Symmetry of gap functions of states that appear in Figs. 1-5. 
For orbital symmetry ($L$),  the commonly used $s$-, $p$-, $d$-wave, 
etc. nomenclature is adopted, with $s^*$, for example, denoting extended 
$s$-wave symmetry.  For spin-singlet states ($S=0$), basis
functions for the scalar order parameter $\Delta_0$ are given, while for 
spin triplets ($S=1$), basis functions for the components of the 
vector order parameter $d_i$ ($i=x,y,z$) are listed.   
In the weak-coupling,
non-relativistic approximation used in this work, all triplet states
with the same {\bf k} dependence are degenerate, independent
of the direction of the {\bf d} vector. 
For instance, 
all 11 $p$-wave states listed for tetragonal crystals in the Sigrist
and Ueda review\cite{Sigrist} ({\bf \^z}$k_x$, {\bf \^x}$k_x\pm 
${\bf \^y}$k_y$, etc.) are degenerate in this approximation 
and correspond to the state labeled here as A. 
Actual gap functions include higher-order Fourier components that
transform in the same way under the symmetry operations of the
crystal (D$_{4h}$) as the basis functions shown in the table.
\label{tab: states}}
\begin{ruledtabular}
\begin{tabular}{lclc}
  & $S$  & $L$ & $\Delta_0({\bf k})$ or  $d_i({\bf k})$ \\
\hline
A &     1 & $p$   & $\sin{\phi}$, $\cos{\phi}$    \\
B &     0 & $s$   & 1      \\
C &     0 & $d$   & $\cos{2\phi}$  \\
D &     0 & $d^*$ & $\cos{2\phi}\cos(k_zc)$  \\
  &     1 & $f$   & $\cos{2\phi}\sin(k_zc)$ \\
E &     0 & $s^*$ & $\cos(2k_zc)$    \\
  &     1 & $p^*$ & $\sin(2k_zc)$ \\
F &     0 & $d$   & $\sin{2\phi}$  \\
G &     0 & $g$   & $\sin{4\phi}$  \\
H &     0 & $d$   & $\sin{\phi}\sin(k_zc)$, $\cos{\phi}\sin(k_zc)$  \\
  &     1 & $p^*$ & $\sin{\phi}\cos(k_zc)$, $\cos{\phi}\cos(k_zc)$\\
I &     0 & $d^*$ & $\cos{2\phi}\cos(3k_zc)$  \\
  &     1 & $f^*$ & $\cos{2\phi}\sin(3k_zc)$ \\
J &     0 & $s^*$ & $\cos(4k_zc)$  \\
  &     1 & $p^*$ & $\sin(4k_zc)$ \\
K &     0 & $d^*$ & $\sin{\phi}\sin(3k_zc)$, $\cos{\phi}\sin(3k_zc)$  \\
  &     1 & $p^*$ & $\sin{\phi}\cos(3k_zc)$, $\cos{\phi}\cos(3k_zc)$ \\
L &     0 & $d^*$ & $\sin{2\phi} \cos(2k_zc)$  \\
  &     1 & $f^*$ & $\sin{2\phi} \sin(2k_zc)$ \\
\end{tabular}
 \end{ruledtabular}
\end{table}

It is interesting that the two-dimensional triplet manifold (A, orange)
emerges despite the
fact that the interactions individually stabilize singlet states.
For this triplet state, the coupling matrix (Eq. \ref{eq: gapeq2}) 
has two degenerate eigenvalues, 
with eigenfunctions $\cos\phi$ and $\sin\phi$,  for each 
component of the vector order parameter ${\bf d}$.    This generates 
the set of allowed $p$-wave states on a cylindrical 
Fermi surface with tetragonal lattice symmetry, as derived, for example,
in Ref. \onlinecite{Sigrist}. Our model corresponds to the
weak-coupling, non-relativistic regime, where
these
states are all degenerate. A chiral member of this set of $p$-wave
triplet states, with ${\bf d} \propto (k_x + i k_y)\hat{\bf z}$, is a 
leading contender for the order parameter in Sr$_2$RuO$_4$.\cite{Mackenzie}
The two-dimensional triplet manifold is stable over a range of 
coupling strength ratios, 
but in all cases, it is limited to the region of the phase 
diagram corresponding to narrow spin interactions.    This can be
understood from the form of the spin interaction.  In the triplet channel,
the spin-fluctuation-mediated interaction is attractive, so to 
induce pairing it must scatter electrons between points on the Fermi surface 
that have order parameters of the same sign.
Wave vectors along the line ${\bf q} = (4\pi/3a, 4\pi/3a, q_z)$, 
for which the spin  interaction is strong, have an in-plane 
component that is slightly larger than $2k_F^\parallel$, 
the diameter of the Fermi cylinder.  Hence, if the spin interaction
is narrowly peaked in the planar directions,  these wave vectors 
contribute little to the total spin coupling. With wider spin
interactions, however, they have a strong pair-breaking
effect if the order parameter has $p$-wave symmetry in the plane. 

\begin{figure}[tbf]
\includegraphics[width=2.0 in]{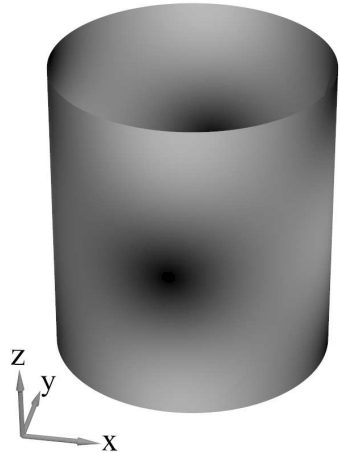}
\caption{Order parameter of the singlet state stable in region D (purple).
White  corresponds to the maximum positive 
value of $\Delta_0$ and black  corresponds to the minimum
negative value. The $\Gamma$ point is at the center of the cylinder.
The full eigenfunction for 
$R=1$, $w_{sp}=0.5\pi/a$, $w_{ph}^\parallel = 0.5\pi/a$, and $\alpha = 1$ 
is plotted, but the lowest-order non-zero Fourier component, 
$\cos{2\phi}\cos(k_zc)$, clearly dominates.
\label{fig: purple}}
\end{figure}

The singlet phase in region D (purple)  has symmetry 
$\cos{2\phi}\cos(k_zc)$
and can be thought of as an extended $d$-wave state, as illustrated
in Fig. \ref{fig: purple}.   It belongs to the  same representation 
as the purely two-dimensional $\cos{2\phi}$ state.    For small $w_{ph}$, 
it is more stable than the $\cos{2\phi}$ state. This is because in
the two-dimensional  $\cos{2\phi}$  state favored by the spin interaction, 
the phonon interaction centered 
at ${\bf q} = (\pi/a, \pi/a, \pi/c)$ is always pair breaking, while the
inclusion of a $\cos(k_zc)$ factor allows the phonon interaction to become
pair forming.  Degenerate with this extended $d$-wave singlet is a set  
of triplet $f$-states, one example of which is ${\bf d} \propto 
\cos{2\phi}\sin(k_zc)\hat{\bf z}$. 
Note that because the two-dimensional spin interaction integrates out, 
the stabilization of the extended $d$-wave singlet and the
degenerate $f$-wave triplets  is 
due entirely to the electron-phonon interaction.

\begin{figure}[tbf]
\includegraphics[width=3.2 in]{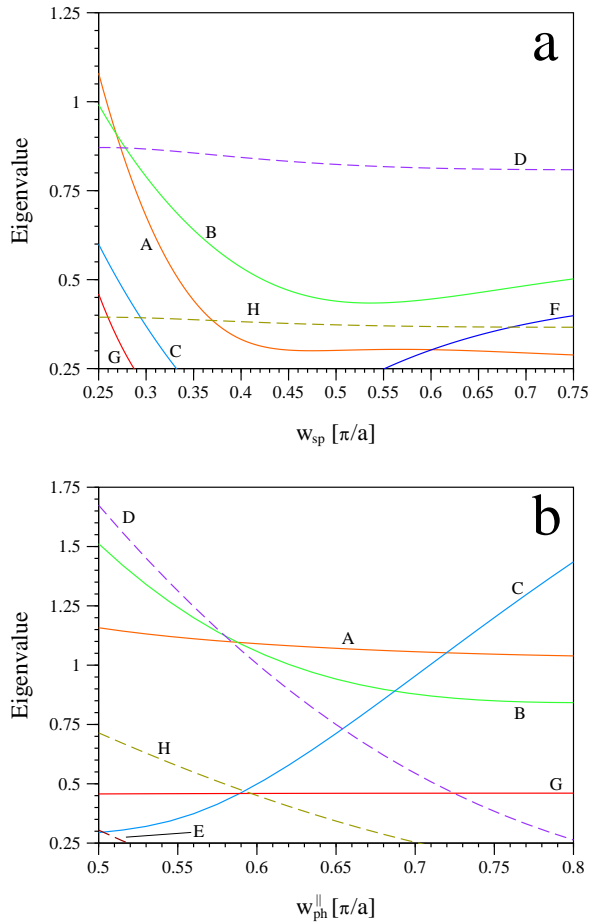}
\caption{Dependence of eigenvalues on (a) width of spin-fluctuation-induced
interaction, $w_{sp}$, and (b) in-plane width of electron-phonon interaction,
$w_{ph}^{\parallel}$, for coupling-strength ratio $R = 1$ and 
electron-phonon anisotropy 
$\alpha = 1$.   In (a), $w_{ph}^\parallel = 0.625\pi/a$, and in (b),
$w_{sp} = 0.25\pi/a$. Dashed lines indicate states with $k_z$-dependent 
eigenfunctions.  See Table \ref{tab: states} for the full symmetry of 
the eigenfunctions.
\label{fig: eigenvalue1}}
\end{figure}

Figure \ref{fig: eigenvalue1} takes vertical and horizontal cuts through 
the phase diagram shown in Fig. \ref{fig: phase1} to illustrate how the 
eigenvalues change with the widths of the phonon and spin interactions. 
In Fig. \ref{fig: eigenvalue1}a, the width of the electron-phonon interaction 
is fixed at $w_{ph}^\parallel = 0.625\pi/a$ with   $\alpha = 1$. 
In Fig. \ref{fig: eigenvalue1}b, the width of the spin-fluctuation-induced 
interaction is fixed at $w_{sp} = 0.25\pi/a$. 
In addition to the four phases that appear in Fig. \ref{fig: phase1}, 
corresponding  to those states with the largest eigenvalues 
within the phase space of the diagram, 
phases with other symmetries appear in Fig. \ref{fig: eigenvalue1}. 
Table \ref{tab: states} lists the symmetry of these states. 
Curve H (gold), for example, corresponds to a manifold of three-dimensional 
singlets with $\Delta_0$ given by linear combinations of 
$\sin{\phi}\sin(k_zc)$ and $\cos{\phi}\sin(k_zc)$, which includes the 
chiral $d$-wave state, $(k_x + ik_y)\sin(k_zc)$ that is compatible
with existing experimental data on the symmetry of the order parameter
in Sr$_2$RuO$_4$.\cite{Zutic}  
When both the spin and phonon interactions are very narrow, or
when both are wide, this three-dimensional $d$-wave singlet (H)
becomes more stable than the two-dimensional $p$-wave triplet (A) that 
is generally believed to be the most likely candidate for the 
pairing symmetry in  Sr$_2$RuO$_4$.\cite{Mackenzie}  
The increased stability of the $d$-wave singlet as compared to 
the $p$-wave triplet in certain parameter regimes arises 
from the same mechanism that was discussed earlier in relation 
to the extended $d$-wave state (D, purple). 
Other symmetries, however, can make even better use of the 
structure of the pairing interaction in these parts of parameter 
space, so the three-dimensional $d$-wave singlet 
does not become the favored phase.

\begin{figure}[tbf]
\includegraphics[width=3.2 in]{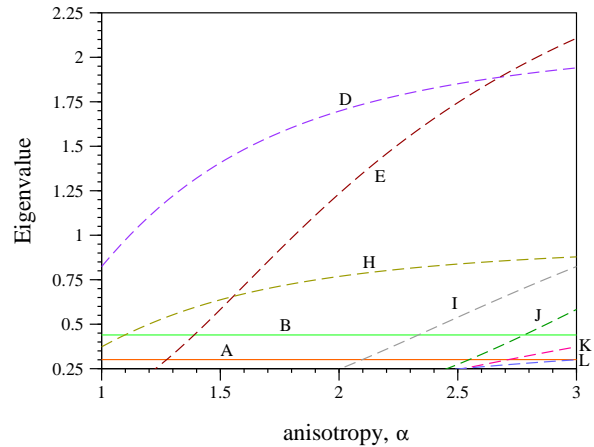}
\caption{Dependence of eigenvalues on the anisotropy of the electron-phonon
interaction, for coupling-strength ratio $R = 1$,
spin-interaction width $w_{sp} = 0.5\pi/a$, and in-plane
phonon-interaction width $w_{ph}^{\parallel} = 0.625\pi/a$.
Dashed lines indicate states with $k_z$-dependent
eigenfunctions.  See Table \ref{tab: states} for the full symmetry of
the eigenfunctions.
\label{fig: eigenvalue2}}
\end{figure}

\begin{figure}[t]
\includegraphics[width=2.7 in]{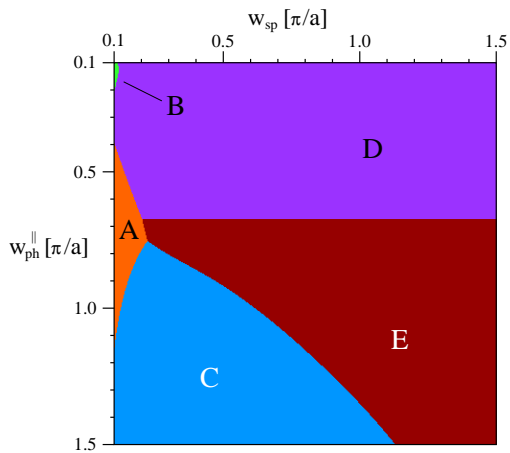}
\caption{Phase diagram for coupling-strength ratio
$R =  1$ and electron-phonon anisotropy $\alpha = 2.5$.
See Table \ref{tab: states} for the symmetry of states that
appear in this phase diagram.
\label{fig: phase2}}
\end{figure}

By narrowing the width of the electron-phonon interaction in the $z$ direction,
the stabilizing mechanism for states with $k_z$-dependent order
parameters can be enhanced.
Figure \ref{fig: eigenvalue2}  illustrates how
the eigenvalues change with anisotropy in the phonon interaction. 
The eigenvalues depend on $\alpha$ only if the order parameter
varies with $k_z$.   As expected, the three-dimensional $d$-wave
singlet (H, gold) is further stabilized as $\alpha$ is increased,
but within our model and within the parameter space explored,
it never emerges as the most stable phase. 
Nevertheless, one can imagine that with a different
structure of electron-phonon coupling in reciprocal space,  this
solution could become stable in some parameter regime.
When $w_{ph}^z$ is very small,  an extended
$s$-wave singlet state (E, dark red) with symmetry $\cos(2k_zc)$ becomes
stable and occupies a large area of the phase diagram (Fig. \ref{fig: phase2}).
As discussed earlier, for states with $k_z$-dependent order
parameters, the two-dimensional 
spin fluctuations integrate out and play no role 
({\it i.e.}, they are neither pairing nor pair breaking). This is reflected 
in Fig.  \ref{fig: phase2} by the fact the the 
phase boundary between region D (purple) and region E (dark red)
is perfectly horizontal.\cite{wspdependence}
Each of those phases represents one singlet (extended $d$ or extended $s$, 
for D and E, respectively) and several triplet ($f$ or extended $p$ for
D and E, respectively) states.
Similar $f$ states have been discussed
in connection with heavy-fermion superconductors, {\it e.g.} in 
UPt$_3$. \cite{upt3}

\section{Conclusions}

The combination of spin- and phonon-mediated pairing 
interactions with different structure in momentum space  can lead to a rich 
variety of unconventional pairing states,  many with
symmetries and parities that could never be stabilized by either
interaction alone.  Within the toy model 
considered here, with a cylindrical Fermi surface, two-dimensional  
spin-fluctuation-mediated pairing, and three-dimensional phonon-mediated 
pairing, there are two-dimensional pairing states with no dispersion
in $z$ and  three-dimensional $k_z$-dependent states.  For the 
two-dimensional states, the interplay of the phonon and spin interactions, 
each of which favors a two-dimensional singlet state,  
can result in a triplet state.
For the three-dimensional states, the symmetry of the model suppresses 
the effect of the spin interaction, and phonon-induced non-$s$-wave states 
become possible.  

This toy model is only loosely connected with the
unconventional superconductor Sr$_2$RuO$_4$, yet it suggests a 
resolution to the seeming paradox regarding the symmetry of the superconducting
state in this compound. While most researchers believe that superconductivity 
is induced by the strong spin fluctuations observed in this system, 
the momentum dependence of these fluctuations, well-known from neutron 
measurements,\cite{spinfluctuation,sfz}
should stabilize the same $d_{x^2-y^2}$ pairing symmetry as in the
cuprates. Yet this symmetry can be excluded based on a variety of 
experiments.\cite{Mackenzie, Nelson}
On the other hand, a large isotope effect on $T_c$ indicates 
the importance of phonons,\cite{isotope} yet 
the $s$ or extended $s$ symmetry favored by phonons 
can be excluded with even more confidence. Our results suggest
the unconventional and rather unusual (in all likelihood, chiral)
symmetry of the superconducting state in Sr$_2$RuO$_4$ emerges
as a result of interplay of electron coupling with both phonons
and spin fluctuations, although neither of these agents separately
can stabilize such a state.   A more quantitative application
of these ideas to Sr$_2$RuO$_4$ requires 
a complete mapping of the electron-phonon coupling in the Brillouin zone 
to complement the existing map of spin fluctuations, as well as consideration
of the full, multi-sheeted Fermi surface.

%============================================================================
\acknowledgments
We would like to acknowledge support from the National Science Foundation,
Grant No. DMR-0210717, and the Office of Naval Research, Grant No. 
N00014-02-1-1046. We also acknowledge useful discussions with D. Agterberg,
M. Braden, and I. Zutic.

%============================================================================


\begin{thebibliography}{99}
\bibitem{notation} In this paper, trigonometric basis functions are used
to label symmetry, so, for example, $d_{x^2-y^2}$ symmetry is labeled 
as $\cos2\phi$, where $\tan\phi = k_y/k_x$.
\bibitem{wspdependence} For three-dimensional eigenfunctions,
the two-dimensional spin-fluctuation interaction cancels out, so 
for fixed $V_{ph}$, the eigenvalue is independent of $V_{sp}$.   
In Figs. \ref{fig: phase1}, \ref{fig: eigenvalue1}, \ref{fig: eigenvalue2}, 
and \ref{fig: phase2}, 
the ratio $R = \lambda_{ph}/\lambda_{sp}$
is constant, so even for fixed $w_{ph}$, the amplitude of $V_{ph}$
varies to keep $R$ constant.  Thus while the eigenvalues
for the three-dimensional eigenfunctions are not constant 
in Fig. \ref{fig: eigenvalue1}a,
they scale in the same way, so the phase line between D and E
in Fig. \ref{fig: phase2} is horizontal.
\end{thebibliography}
\end{document}